\documentclass[10pt,journal,compsoc]{IEEEtran}
\usepackage{tikz}
\usepackage{amsmath}

\usepackage{amsmath}
\usepackage{soul}
\usepackage{algorithm}
\usepackage{algorithmic}
\usepackage{multirow}
\usepackage{threeparttable}
\usepackage{amsmath,amssymb,amsfonts}
\usepackage{algorithmic}
\usepackage{graphicx}
\usepackage{textcomp}
\usepackage{xcolor}
\usepackage{listings}
\usepackage{url}
\usepackage{verbatimbox}
\usepackage{colortbl}
\usepackage{subfigure}
\usepackage{ragged2e}

\ifCLASSOPTIONcompsoc
  \usepackage[nocompress]{cite}
\else
  \usepackage{cite}
\fi
\ifCLASSINFOpdf
\else
\fi
\begin{document}
%
\title{SIAT: A Systematic Inter-Component Communication Analysis Technology for Detecting Threats on Android}
%
%
%
%

\author{Yupeng~Hu,~\IEEEmembership{Senior Member,~IEEE,}
        Zhe~Jin,
        Wenjia~Li,~\IEEEmembership{Senior Member,~IEEE,}
        Yang~Xiang,~\IEEEmembership{Fellow,~IEEE,}
        Jiliang~Zhang,~\IEEEmembership{Senior Member,~IEEE}
\IEEEcompsocitemizethanks{
\IEEEcompsocthanksitem Jiliang Zhang the corresponding author.
\IEEEcompsocthanksitem Y. Hu, Z. Jin and J. Zhang are with the College of Computer Science and Electronic Engineering, Hunan University, Changsha 410082, P.R. China. E-mail: \{yphu, jinzhe, zhangjiliang\}@hnu.edu.cn.

\IEEEcompsocthanksitem W. Li is with the Department of Computer Science, New York Institute of Technology,
New York 10023, USA. E-mail: wli20@nyit.edu.

\IEEEcompsocthanksitem Y. Xiang is with Digital Research and Innovation Capability Platform, Swinburne University of Technology, Australia. E-mail:yxiang@swin.edu.au.

}
\thanks{Manuscript received June 5, 2020.}}

%
%

\markboth{Journal of \LaTeX\ Class Files,~Vol.~14, No.~8, August~2015}%
{Shell \MakeLowercase{\textit{et al.}}: Bare Demo of IEEEtran.cls for Computer Society Journals}
%




\IEEEtitleabstractindextext{%
\justifying
\begin{abstract}
In this paper, we present the design and implementation of a Systematic Inter-Component Communication Analysis Technology (SIAT) consisting of two key modules: \emph{Monitor} and \emph{Analyzer}. As an extension to the Android operating system at framework layer, the \emph{Monitor} makes the first attempt to revise the taint tag approach named TaintDroid both at method-level and file-level, to migrate it to the app-pair ICC paths identification through systemwide tracing and analysis of taint in intent both at the data flow and control flow. By taking over the taint logs offered by the \emph{Monitor}, the \emph{Analyzer} can build the accurate and integrated ICC models adopted to identify the specific threat models with the detection algorithms and predefined rules. Meanwhile, we employ the models' deflation technology to improve the efficiency of the \emph{Analyzer}. We implement the SIAT with Android Open Source Project and evaluate its performance through extensive experiments on well-known datasets and real-world apps. The experimental results show that, compared to  state-of-the-art approaches, the SIAT can achieve about 25\%$\sim$200\% accuracy improvements with 1.0 precision and 0.98 recall at the cost of negligible runtime overhead. Moreover, the SIAT can identify two undisclosed cases of bypassing that prior technologies cannot detect and quite a few malicious ICC threats in real-world apps with lots of downloads on the Google Play market.
\end{abstract}

}

\maketitle

\IEEEdisplaynontitleabstractindextext

%
\IEEEpeerreviewmaketitle

\IEEEraisesectionheading{
\section{Introduction}\label{introduction}}

%
%
%
%


%
\IEEEpeerreviewmaketitle
\IEEEPARstart{I}{n} recent years, we have witnessed explosive growth in the number of mobile devices, and the large quantity of diversified mobile applications (apps) on those mobile devices have made our daily lives much more convenient and enjoyable. However, with the rapid growth of mobile apps, they have increasingly become the target of mobile malware authors, who generally develop and distribute mobile malware that aims at stealing and disclosing various types of sensitive and valuable information that is associated with either mobile user or device, such as credit card number, real-time geolocation, and International Mobile Equipment Identity (IMEI) number, etc. Malware has become one of the most significant security threats to mobile operating systems, especially Android.

In the Android system, the widely used Inter-Component Communication (ICC)\cite{understanding} plays an essential role between the components of apps that are isolated in different sandboxes. Apps pass messages between each other by passing the \textit{intents}, which are passive data structures holding the abstract descriptions of operations to be performed between components. Such a flexible method contributes a lot to functionality reuse and data sharing; however, it also exposes a vulnerable surface to several security attacks. In the context of ICC mechanism scenarios, apps whose developers overlooked security issues often suffer from risky vulnerabilities such as intent hijacking and spoofing\cite{chin}, resulting in sensitive user data leak or privilege misuse by other apps, particularly mobile malware. Besides, two or more malicious apps with ICC paths could even collude on stealthy attacks that neither of them could accomplish alone\cite{privilege}. In these attacks, malicious apps send and receive intents in a way that looks like as if those are ordinary message exchanges. By this means, they can often easily bypass those classical malware detection approaches which regularly inspect apps individually.

Many existing ICC-relative research works \cite{epicc}\cite{chex} focus on detecting vulnerabilities in benign apps. None of these techniques could identify ICC paths with attack behaviors. Recently, most of the research works that aim at identifying ICC paths with attack behaviors are in two categories: static analysis and running protection. A static analysis  approach often extracts sensitive ICC paths by matching attributes and tracking data flow (e.g., IC3\cite{ic3}, AmanDroid\cite{amandroid}, DIALDroid\cite{dialdroid}). However, even the state-of-the-art static analysis based approaches suffer from a large number of false positives, the reason being that they could not validate the data format through static analysis when facing the reflection and unreachable code. As a result, ignoring the validation of the data format in the static analysis will lead to an ICC path that does not occur in reality. Alternatively, runtime protection based approaches, e.g., XManDroid\cite{xmandroid} and SCLib\cite{sclib}, either enforce mandatory access control according to the predefined policy set or ask about the end-user's decision for access permission to protect them from attacks when apps communicate with each other using the ICC mechanism. However, those runtime protection based approaches only pay attention to the information acquired before receiving intent, so that ignores actual behaviors of the receiver. Particularly, they determine whether to prohibit the mobile app from receiving intent according to various information (such as app permissions, intent attributes, and intent filter attributes), which makes these runtime protection techniques unable to provide a solution for accurately identifying ICC paths with attack behaviors.

To overcome the shortcomings of the existing ICC-based malware detection approaches, we propose a Systematic ICC Analysis Technology (SIAT), which identifies ICC paths with attack behaviors through analyzing the monitor information of components and data flows in the ICC processes at runtime. SIAT consists of two key modules: \emph{Monitor} and \emph{Analyzer}. The \emph{Monitor} recognizes the application information exchanged by the transmission intent at runtime, and closely monitors the spread of sensitive information in the intent transmission process through dynamic taint analysis. Once the \emph{Analyzer} obtains the information from the \emph{Monitor}, it establishes \textbf{\emph{ICC threat models}} (we describe the details in Section \ref{threat model}) and identifies the ICC paths with attack behaviors based on these threat models. The SIAT is able to deal with three typical ICC related security threats, namely intent hijacking, intent spoofing, and especially intent collusion attack.

Our key contributions are summarized as follows:

\begin{itemize}

  \item We propose the SIAT\footnote{https://github.com/JinxKing/SIAT.}, a more accurate systematic approach for identifying the malicious ICCs between apps by migrating the taint tag approach named TaintDroid \cite{taintdroid} to trace and analyze sensitive intent data, depending on its two key modules, namely \emph{Monitor} and \emph{Analyzer}.

   \item \emph{Monitor}, an Android framework extension that monitors intent transfers and data flows in ICC at runtime through systemwide tracing and accurate analysis of tainted data. In particular, the SIAT redefines the service primitives and sensitive data for intent, and more than eighty taint tags for tracing the sensitive intent communications by revising the TaintDroid at method-level and file-level.

   \item \emph{Analyzer}, an approach for building threat models by taking over the taint logs offered by the \emph{Monitor} in seamlessly to identify the specific threat models (i.e., intent hijacking, spoofing, and collusion attack) with the identification algorithms and predefined rules.

   \item We have evaluated the performance of the SIAT through extensive experiments on both malwares and benign apps composed of several well-known datasets and thousands of real-world Android apps.

\end{itemize}

The remainder of this paper is organized as follows. Related works are discussed in Section \ref{related work}. Section \ref{threat model} discusses the threat models SIAT focuses on. Section \ref{SIAT} describes the overall architecture of SIAT. Section \ref{evaluation} presents the comprehensive performance evaluation that we have conducted for SIAT. Before drawing conclusion in Section \ref{conclusion}, we introduce the in depth discussion in Section \ref{discussion}.

\section{Related Work}
\label{related work}
In this section, we review some of the related works that are the starting point of our research. Inter-app threat issues have attracted a lot of research efforts. The state-of-the-art approaches could broadly be divided into two categories: single-app analysis and app-pair analysis. And the analysis approaches of each category also are in two types: static analysis and dynamic (a.k.a., runtime) analysis.

\textbf{Single-app analysis}. The static single-app analysis approaches, such as \cite{chex,amandroid,FlowDroid,IccTA,ICCDetector}. CHEX\cite{chex}, can identify the component hijacking vulnerabilities through static data flow analysis. Amandroid\cite{amandroid} focuses on analyzing inter-component data flows and track the interaction of the components.  FlowDroid\cite{FlowDroid} improves the precision for static taint analysis between components in an application. ICCTA\cite{IccTA} focuses on privacy leakage based on the static taint analysis. ICCDetector\cite{ICCDetector} builds a model to detect malwares via extracting the ICC features and training with a set of benign apps and malwares.

The dynamic single-app analysis approaches \cite{taintdroid,IntentFuzzer,intentdroid,dazed} monitor the app at runtime. As a data flow tracing method, TaintDroid\cite{taintdroid} monitors the system at runtime and tracks the taint transmission so as to detect the privacy leakage. IntentFuzzer\cite{IntentFuzzer} identifies the vulnerable interfaces by dynamically sending test intents to the components. IntentDroid
\cite{intentdroid} tests eight different vulnerabilities caused by unsafe handing of coming ICC intent data. DazeDroid\cite{dazed} fully-automated extracts the components and can fuzzing all interfaces in apps.

\textbf{App-pair analysis.} In the static app-pair analysis technologies, ApkCombiner \cite{ApkCombiner} directly combines two apps into a single app and uses the single static data flow analysis method to identify sensitive ICC methods. Both Epicc \cite{epicc} and IC3 \cite{ic3} extract and analyze the information from apps. COVERT \cite{covert} employs a compositional analysis method for finding inter-app vulnerabilities. JITANA\cite{jitana} can analyze multiple android apps simultaneously. DidFail\cite{didfail} is designed to detect the information leakage between activities. DIALDroid\cite{dialdroid} analyzes each app and adopts the database to calculate the sensitive ICC path. MRDroid\cite{mrdroid} builds a component interaction map based on MapReduce, and assesses the inter-app threats of apps with the ICC map.

Most of the dynamic app-pair analysis technologies enforces security policies only at the sender so as to protect users from inter-app threats. XManDroid \cite{xmandroid} is the first approach proposed to prevent application-level privilege escalation attacks through enforcing the permission policies. FlaskDroid \cite{flaskdroid} provides mandatory access control strategy simultaneously for both Android's middleware and kernel layers to prevent privilege escalation and collusive data leak. SCLib \cite{sclib} proposes an approach that performs inter-app mandatory access control for defending against component hijacking without modifying the Android system. SPEAR \cite{separ} and SEALANT \cite{sealant} combine the static analysis and enforcing security policy to provide end-user protection.

\textit{Different from existing dynamic ICCs detection technologies, SIAT, not only inspects the sender’s but the receiver’s intent related behaviors by migrating the dynamic single-app analysis approach TaintDroid to the systemwide tracing of intent data among multiple apps/components, to improve the accuracy of threats detection significantly.}

\begin{figure}[t]
\centering
\includegraphics[scale=0.7]{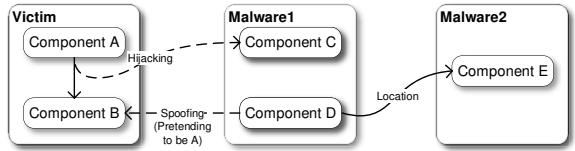}
\caption{The intent hijacking, spoofing and collusion attack models that bring forth malicious ICC paths to be identified.}
\label{threatfig}
\end{figure}

\section{Threat Models}

\label{threat model}
Communications between components of mobile application (a.k.a., app) are achieved via sending and receiving intents, which are generally used with methods to invoke activity, service, and broadcast receivers. In this section, we present the generation mechanism of the malicious ICC paths that are incurred by malwares with intent in three typical threat models, i.e., intent hijacking, intent spoofing, and intent collusion attack. As shown in Figure \ref{threatfig}, the following threat models that SIAT focuses on may lead to different malicious ICC attack behaviors, such as privilege escalation, sensitive data leak, and so on. \textit{It is worth noting that SIAT mainly focuses on identifying the intent threat model but not the specific attack behaviors.}


\textbf{Intent hijacking}. As depicted in Figure \ref{threatfig}, in intent hijacking, the implicit intent may never reach the expected component, but it is instead intercepted by an unauthorized app. In \textit{Victim} app, when the \textit{Component A} sends intent to \textit{Component B}, the \textit{Malware}1 app can obtain the intent by setting the attributes matched with the intent in the intent-filter of the component. As a result, it is easy to cause data leakage during the process of intent hijacking. If the data (e.g., location, contacts) requires permission to obtain, and the intent does not restrict the receiver with permission, \textit{Malware}1 obtains the sensitive data without the necessary permission. In this case, the \textit{Malware}1 escalates the privilege by hijacking the intent besides stealing the sensitive data.

\textbf{Intent spoofing}. Figure \ref{threatfig} illustrates how intent spoofing works. If the \textit{Victim} app discloses that the \textit{Component B} expects to receive intent from the \textit{Component A} or some other components, and it has no proper restrictions on attributes, the \textit{Malware}1 may then pretend to be \textit{Component A} and send intent to the \textit{Component B}, which could trigger the corresponding action of the \textit{Component B}.


\textbf{Intent collusion attack}. Intent collusion attack generally refers to the situation in which two apps work together to accomplish malicious behaviors that a single app cannot achieve solely by itself. As Figure \ref{threatfig} shows, \textit{Component D} in \textit{Malware1} sends an intent with location data to \textit{Malware2}. Afterwards, \textit{Component E} receives the intent and then sends out the SMS message with location data. Since all these actions can be performed in the background, it is difficult for users to perceive. To make things worse, it is very difficult to identify each app as a malware if merely inspecting its own behavior separately. During the process of intent collusion attack, the malicious apps pretend to communicate normally with each other, which brings great challenges to identify those apps successfully as malwares.


%
%

\vspace{-0.0in}
\textbf{Coincidental malicious ICCs.} Furthermore, not only do we need to distinguish the aforementioned malicious ICCs from ordinary application communications, \textit{but we also need to differentiate the malicious ICC with `coincidental malicious ICC', which is omitted or lacks accurate analysis in the existing research efforts}. If some developers do not have strict control over the attributes of the intent or intent-filter, which may lead to an accidental match between the apps, then when an app receives data from the intent, it may find out that the data do not meet the requirements of the intent, which will cause the app to stop execution. Especially if the ICC path is incorrectly considered to involve sensitive APIs, it will be classified as a sensitive ICC path and thus be regarded as an initiative attack path launching by a malware, in this situation, the ICC path is called a `coincidental malicious ICC'. Ignoring that can result in high false positive rates. We showcase the solution of this issue in Section \ref{subsec:RQ2}.

\section{The SIAT}
\label{SIAT}

\begin{figure}[tbp]
\centerline{\includegraphics[scale=0.7]{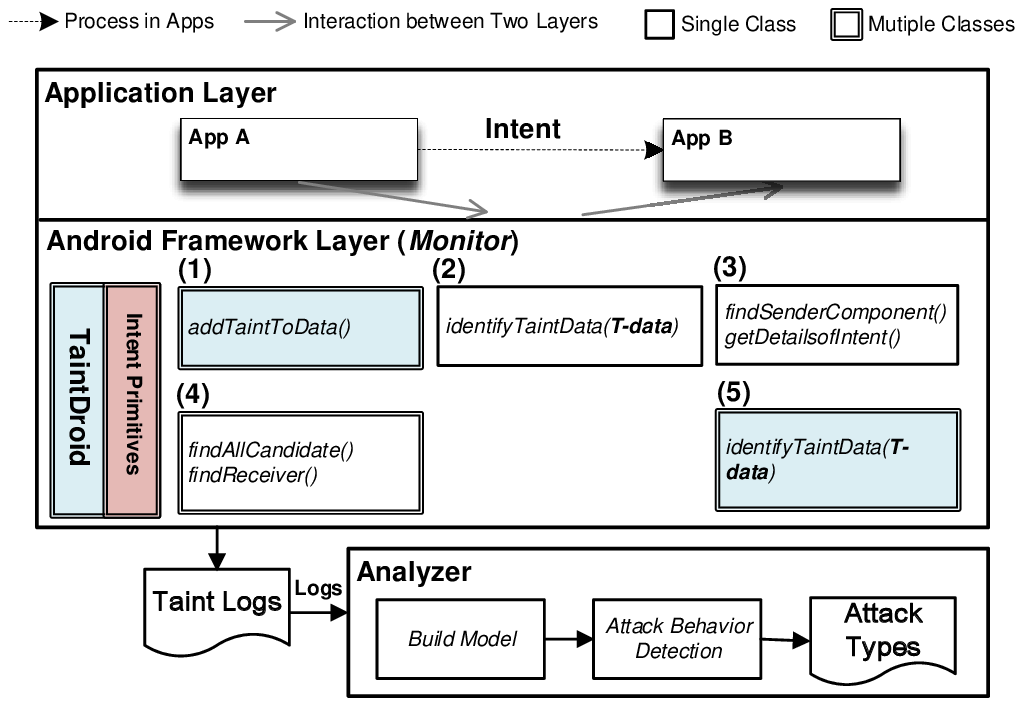}}
\caption{The architecture of SIAT. The five functions cooperate with TaintDroid to trace the intent via the intent primitives.}
\label{arcoverview}
\end{figure}

\begin{figure*}[htbp]
\centerline{\includegraphics[scale=0.8]{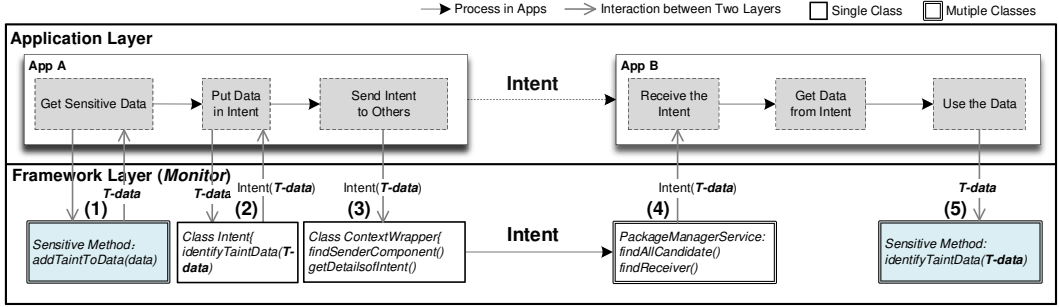}}
\caption{The overall workflow (data flow and control flow) of \emph{Monitor} in inter-component communication.}
\label{monitoroverview}
\end{figure*}

This section introduces the SIAT, which can identify the malicious ICCs with attack behaviors by analyzing the real-time intent information and its data and control flows.

\subsection{Architecture Overview}
\label{subsec:arch-detail}
As a starting point, Figure \ref{arcoverview} presents the architecture of the proposed SIAT, which consists of two key modules: \emph{Monitor} and \emph{Analyzer}. By analyzing the data tainted with tailored TaintDroid \cite{taintdroid} and redefined intent service primitives, the SIAT provides a real-time systematic tracing of privacy-sensitive data and attack visibility into how the collaborative malicious behaviors take place via intent.

The \emph{Monitor}'s main contributions lie in the migration of Taintdroid to ICC attack identification not only at the data flow but at the control flow. Relying on the main five functions, as shown in Figure \ref{arcoverview}, the \emph{Monitor} extends the framework layer of the current Android operating system based on the Android Open Source Project. By cooperate with core methods of TaintDroid via intent primitives, the \emph{Monitor} inspects the relevant components on an ICC path and the data and control flow associated with the intent at runtime, and further identifies the intent's sender, the intent-matched component, and the receiver at several critical points of the ICC process. In Figure \ref{arcoverview}, the single and multiple classes  denote the number of classes involved in the implementations of functions of \emph{Monitor}. The overall workflow of \emph{Monitor} is listed as follows:

\textbf{(1) Set Taint.} When the sender gets sensitive data from sensitive source, using the function \verb+addTaintToData()+, the \emph{Monitor} taints the sensitive data (e.g., location, phone number) and adds a variable tag (an 8-bit taint number) to it, which clearly labels its source. In this paper, we name the sensitive data tainted as \textit{T-data}.

\textbf{(2) Check Intent.} When the sender sets the intent attributes (e.g. extra, action), the \emph{Monitor} checks the \emph{T-data} to see whether or not it is tainted or retainted through the function \verb+identifyTaintData(T-data)+.

\textbf{(3) Sending Intent.} If there is a sender that calls the system API to send an intent, the \emph{Monitor} identifies the identity of the sender and the details of the intent using functions \verb+findSenderComponent()+ and \verb+getDetailsofIntent()+.

\textbf{(4) Receiving Intent.} Upon obtaining the best matched component, the \emph{Monitor} is able to find out all candidates as well as the real receiver of intent by means of the functions \verb+findAllCandidate()+ and \verb+findReceiver()+.

\textbf{(5) Check Taint.} When the receiver extracts the \emph{T-data} from an intent, as long as the it calls the sensitive APIs, the \emph{Monitor} will check if any parameter in the APIs is \emph{T-data} so as to identify the source of the data by utilizing the function \verb+identifyTaintData()+.

As depicted in Figure \ref{arcoverview}, the \emph{Analyzer} needs to analyze the \emph{APK} (the abbreviation of AndroidPackage) files and obtain the app's attributes and component information, to build a complete ICC model with the taint logs generated by the \emph{Monitor}. According to the matching results of the ICC information with the predefined rules, the \emph{Analyzer} then determines whether there is any aforementioned ICC threat model or not, so that we can distinguish malwares from benign apps. More details regarding the \emph{Monitor} and the \emph{Analyzer} are presented in Section \ref{subsec:monitor} and \ref{subsec:analyzer}.

\subsection{Monitor}
\label{subsec:monitor}

\subsubsection{TaintDroid Migration}
To best suit our requirements, we need to migrate the taint tag method TaintDroid to trace sensitive intent data not only at the data flow but at the control flow, as shown in Figure \ref{monitoroverview}. Thus we revised the TaintDroid for inspecting the specific data flow as well as control flow of sensitive data in the ICC process, so that the \textit{Monitor} can trace the flow of privacy-sensitive data at runtime, which basically meets our requirements of taint analysis. In particular, the SIAT defines a set of intent service primitives and more than eighty taint tags for analyzing the sensitive intent data flows by revising the TaintDroid at method-level as well as file-level as follows.

Firstly, as highlighted in Figure \ref{arcoverview}, we have defined a set of service primitives for intent communications, which encapsulate the sensitive data operation functions and works as a middleware between the core methods of TaintDroid and the Android intent mechanism by method-level and file-level revisions. The new intent primitives encapsulate the functions of returning the source of current taint, obtaining the next tag with original taint, setting/getting the tags, and so on. In this way, the main five functions at the framework layer shown in Figure \ref{arcoverview} are able to cooperate with TaintDroid efficiently. For instance, when apps call APIs to get those privacy-sensitive data, based on the function \verb+addTaintToData(data)+ in Figure \ref{monitoroverview}, we taint the data as the \emph{T-data} by which we can trace and distinguish the data from others. Also, we can extract the tag from \emph{T-data} and identified the \emph{T-data} by comparing the number with the function \verb+identifyTaintData(T-data)+. The bit vector of tag is null if the data is not tainted.

Secondly, by revising the main files of TaintDroid, our \textit{Monitor} defines a group of new sensitive data and eighty taint tags for identifying them in intent communications. For example, the sensitive location data, \verb+TAINT_LOCATION_Latitude=0x00010004+, \verb+TAINT_LOCATION_Longitude=0x00010008+.
Not only do we consider the privacy-sensitive data (e.g., locations, contacts, phone state) as sensitive, but we also regard the information that the user inputs or acquires from other files, other content providers (e.g., \verb+Shareference+,)\footnote{Shareference is a persistent storage method provided by Android.} as sensitive.

Afterward, to inspect the sensitive data in interested APIs, we take advantage of a machine-learning technology for achieving the most likely source and sink methods \cite{rasthofer2014machine}.

\subsubsection{Monitor Implementation}
Figure \ref{monitoroverview} demonstrates the specific work mechanism of \emph{Monitor} in ICC process. As mentioned before, the main task of \textit{Monitor} is to identify the sender, the receiver, and the data flow between them. To monitor the data flow and control flow in the ICC process, the two key challenges the \emph{Monitor} have to deal with are: firstly, where the data in the intent by the sender initially comes from; secondly, where the data in intent finally goes to in the receiver. In this regard, the data is tagged as tainted if it comes from a privacy-sensitive source. All possible ways the information could leave the device is represented as \emph{Sink}. If the tainted data is found in the \emph{Sink}, there may be an inevitable leakage of the privacy-sensitive data. The main challenge of designing the \textit{Monitor} lies in the accurate identification and recording of the delivery chain of the privacy-sensitive data at runtime among multiple apps. We describe the design and implementation of \emph{Monitor} by answering the following four questions:

\textbf{Is there any sensitive data in the intent?} When apps provide data for an intent, based on the \textit{T-data}, we inspect the parameters of the data to see if it has been tainted with the function \verb+identifyTaintData(T-data)+. If so, the tainted data will be retained with a new variable tag and a source code to show the source of the intent clearly. If not, the data will be tainted and marked as from this specific intent. We have defined more than eighty types of tags to identify the sensitive data, e.g., \verb+TAINT_sharepreference=0x00010018+, \verb+TAINT_network_state=0x00010012+. In this way, the \emph{Analyzer} can easily figure out where the sensitive data comes from in the receiver at runtime.

\textbf{Who is the sender of the intent?} When an app sends an intent, we need to capture the sending event and the information of the source component. The operation of calling API (e.g., \verb+startService(intent)+) for sending an intent is generally inherited from another class for \verb+Activity+ or \verb+Service+ component. The \verb+BroadcastReceiver+ will execute the calling operation by acquiring the \verb+Context+ object and using the API in \verb+Context+. As Figure \ref{monitoroverview} shows, in practice, the implementation of these APIs is in the \verb+ContextWrapper+ class. Therefore, by integrating the codes for the functions of \verb+findSenderComponent()+ and \verb+getDetailsofIntent()+ into the \verb+ContextWrapper+ class, we will immediately notice whenever an intent is sent. Then, we can utilize the Java refection method to figure out which component calls the API to send the intent and which package the component belongs to.

\textbf{Who is the receiver of the intent?} After capturing the sending event, we want to know which component becomes the candidate as its attributes match with those of the intent, and which component receives the intent at the end. In our design, we integrate two functions \verb+findAllCandidate()+ and \verb+findReceiver()+ into the \verb+PackageManageService+ (PMS) for querying the components that match with the intent by traversing the components of all apps as candidates. There are three types of components involved in the intent matching: first, for the receiving component of \verb+Activity+, if there are more than one matched components, the PMS selects one component from the list of candidates through comparing their priorities, such as the preferred order and so on. Alternatively, the PMS can also ask the user to choose one component; secondly, for the receiving component of \verb+Service+, the \textit{Monitor} will choose the first candidate; thirdly, for the receiving component of \verb+broadcast receiver+, the PMS sends intent to all candidates. Therefore we can monitor all candidates and the actual receiver of intent in PMS.

\textbf{How does the receiver use the sensitive data extracted from intent?} The \emph{Monitor} inspects the data outputted to a file or sent to another device so as to determine whether it is tainted. If it is, the \emph{Monitor} identifies the source of the data through the tag in the\emph{T-data} with function \verb+identifyTaintData(T-data)+. Therefore, it can indicate the way how the receiver uses the data extracted from intent. Also, we take more sensitive methods into consideration in the \emph{Monitor}, such as the methods to store data in \verb+Shareference+ and database (e.g.,  \verb+Editor.putString()+). It is worth noting that these sensitive methods don't need to apply for permissions; thus they could easily be overlooked.

\subsection{Analyzer}
\label{subsec:analyzer}

The \emph{Analyzer} takes over the taint logs outputted by the \emph{Monitor} in a seamless way to build the ICC models and further identify the specific threat models by matching with the identification algorithm and predefined rules in the following two essential parts.

\subsubsection{Model Building}
As Table \ref{table:IACmodel} depicts, a threat model to be built in \emph{Analyzer} is composed of three objects, including the \emph{Sender}, the \emph{Intent}, and the \emph{Receiver}. To ensure  efficiency, we only adopt the most useful attributes, e.g., the \textit{taint data}, which denotes the new sensitive data for intent. Based on Table \ref{table:IACmodel}, there are two key technologies below for building models:

\textbf{Intent data extraction.} To build accurate threat models, the \emph{Analyzer} needs to extract the intent related information from logs. Firstly, the \emph{Analyzer} needs to extract information such as package names and permissions of each app by analyzing the \emph{APK} files. The package names enable the \emph{Analyzer} to obtain the \emph{process ID} of the app, which is a unique identifier assigned to each app by the Android system. Every log contains a \emph{process ID} that is associated with the app, which generates the log. Secondly, based on the \emph{process ID}, the \emph{Analyzer} filters all uncorrelated logs that are from the Android system itself and the other apps, and only retains the logs regarding the apps that we are interested in. Thirdly, the \emph{Analyzer} reads every filtered log and extracts the intent relevant information that is useful for building the ICC models. According to the information extracted from the logs, as shown in Table \ref{table:IACmodel}, the \emph{Analyzer} can straightly acquire the attributes in intent, the \emph{source methods} in the sender, and the \emph{startCompt} and \emph{sink methods} in the receiver.

Furthermore, the \emph{Analyzer} needs extra work to analyze the attributes related to the permissions in the models by identifying the attribute \emph{permissions required} in the sender based on the permissions required to generate the tainted data in the intent. The attribute \emph{permissions required} in the receiver is adopted to implement the \emph{sink method}. Hence, for the sender, the attribute \emph{permissions lacked} denotes the one that the sender doesn't have, but the receiver requires, and vice versa, for the receiver.

\begin{table}[t]
\footnotesize
\caption{The most useful attributes of application and intent adopted in our ICC model analysis.}
\label{table:IACmodel}
\begin{center}

\begin{threeparttable}
	\begin{tabular}{ c | c | c }
		\hline
		\textbf{Sender}&\textbf{Intent}&\textbf{Receiver}\\
		\hline\hline
		process ID&action&process ID\\
		package& categories & package\\
		components& type & components\\
		permissions required & scheme & permissions required\\
		permissions lacked& taint data1 & permissions lacked\\
		source methods\tnote{1} & $\cdot\cdot\cdot$ &  sink methods \\
		candidates& taint data\emph{N} & startCompt\\
	
		\hline
	\end{tabular}
	\begin{tablenotes}
        \footnotesize
        \item[1] The sensitive method where the tainted data comes from.
      \end{tablenotes}
\end{threeparttable}
\end{center}
\end{table}

\textbf{Models deflation.} After extracting the information required for detecting ICC process, the \emph{Analyzer} is able to build the ICC models in an efficient first-in-first-out way so as to stay consistent with the Android intent mechanism. Firstly, the \emph{Analyzer} starts the building process with the logs generated by the APIs called for sending an intent. After that, whenever the \emph{Analyzer} detecting the other log data comes from the APIs called for sending an intent, it will restart to build a new one after building the current ICC model successfully. While this straightforward design leads to a models inflation challenge we have to deal with due to the following two reasons:

\begin{figure}[thb]
\centerline{\includegraphics[scale=0.6]{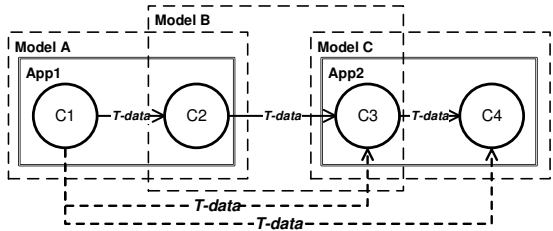}}
\caption{The threat models of multiple components.}
\label{muti-app-model}
\end{figure}

Firstly, there are redundant models generated by the multi-hop intent transfers between multiple components, as shown in Figure \ref{muti-app-model}. Figure \ref{muti-app-model} depicts the generation of redundant models (\emph{Model A, B and C}) built by the process mentioned above based on the four components in a streamlined way. It is incorrect in that the source or destination of the sensitive data tainted as \emph{T-data} is not the real component who sends or accepts the intent. In \emph{Model B} and \emph{Model C}, the \emph{T-data}'s source is considered as \emph{C}2 and \emph{C}3 respectively, however, the real source and destination is \emph{C}1 of \emph{Model A} and \emph{C}4 of \emph{Model C} respectively.

Secondly, there are extra models generated by the Android for launching the internal components that we are not concerning. For instance, if the destination component is \verb+activity+ and there are more than one matching with the intent attributes, the Android will deliver the intent named \emph{a} to the \verb+ResolverActivity+ firstly to let the user choose the desired component. After the user selecting the destination component, instead of the sender, the \verb+ResolverActivity+ then transfers a new intent named \emph{b} to the destination.

To address the models' inflation issue, the \emph{Analyzer} takes advantage of a deflation technology to eliminate the redundant models as follows. The deflation technology can build an ordered model list based on the components in the sender and the receiver of a single model. It then traverses each model to compare the taint tag for identifying the real source in the sender and the destination/sink in the receiver, respectively. In this way, the three models in Figure \ref{muti-app-model} will be condensed into one, and the \emph{Analyzer} is able to figure out if the final receiver $C$4 starts a private component to leak out the sensitive data after receiving the intent. Similarly, the proposed deflation technology also can remove the unnecessary and interfering models come from the Android system's internal components. For the example mentioned above of \verb+ResolverActivity+ in the second reason, using the deflation technology, the \emph{Analyzer} is able to simplify the two models in the intent transfer process as one by executing the following steps: replacing the sender of intent \emph{b} with the sender of \emph{a} and then keeping the intent \emph{b} meanwhile discarding the \emph{a}.

\begin{algorithm}[t]
\label{algorithm:attack}
\caption{Threat models identification}
\renewcommand{\algorithmicrequire}{\textbf{Input:}}	\renewcommand{\algorithmicensure}{\textbf{Output:}}
\begin{algorithmic}[1]
\REQUIRE $models \Leftarrow$ all models
\ENSURE $ModelTypes \Leftarrow$ a map of model and type
\STATE Let $model$ be a model in $models$.
\STATE Let $sender$ be the sender object in a model.
\STATE Let $intent$ be the intent object in a model.
\STATE Let $rcver$ be the receiver object in a model.
    \FOR{each $model \in models$}
    \FOR{each $component \in sender.components$}
    \IF{$component \in Intent.candidates$}
    \STATE $add$ $(model,``hijacking")$ to $ModelTypes$
    \STATE $continue$
    \ENDIF
    \ENDFOR
    \IF {$(rcver.taintleak=true)\wedge(sender.lackpms=null)$}
        \STATE $add$ $(model,``hijacking")$ to $ModelTypes$
    \ELSIF{$(sender.lackpms\not=null)\wedge(rcver.lackpms=null)$}
        \STATE $add$ $(model,``spoofing")$ to $ModelTypes$
    \ELSIF{$(rcver.startCompt=true)\wedge(rcver.lackpms=null)$}
        \STATE $add$ $(model,``spoofing")$ to $ModelTypes$
    \ELSIF{($sender.lackpms\not=null)\wedge(rcver.lackpms\not=null)$}
        \STATE $add$ $(model,``collusion")$ to $ModelTypes$
    \ENDIF
    \ENDFOR

\end{algorithmic}
\end{algorithm}

\subsubsection{Threat Models Identification}
The \emph{Analyzer} implements Algorithm 1 to identify the possible threat models in the ICC models after the previous steps for building all models. According to the attributes in Table \ref{table:IACmodel}, Algorithm 1 considers five different cases, which cover all attack types we target in this paper and iterates through each model to find the best matching case.

\textbf{Case 1} (line 6-11): Algorithm 1 traverses all components in the \emph{sender} app to examine carefully if the list of candidates contains a component from the sender when the receiver component does not belong to the sender. If so, we add the model and the attack type ``hijacking" to \verb+ModelTypes+. In this case, we deem that the instance illustrated in Figure \ref{threatfig} is happening and the candidate component from the sender should be the real destination instead of the receiver component. Therefore, the receiver component hijacks the intent, which is supposed to be sent to another component.

\textbf{Case 2} (line 12-13): If there is some data from the intent used by a specific sensitive method in the receiver while the sender has the permissions related to sensitive method in the receiver, we add the model and the attack type ``\emph{hijacking}" to \verb+ModelTypes+. Afterward, the data extracted from the intent will be utilized by the sensitive method in the receiver, which means the receiver proactively acquires the private data from the sender through obtaining the intent. The situation that the sender lacks the permission related to sensitive method in the receiver is considered as illegal behavior and is classified as another type of threat.

\textbf{Case 3} (line 14-15): When the sender lacks the permission that the receiver needs for calling a sensitive method in the ICC process, but the receiver does have the permission for the data from the sender, we determine that the sender sends the intent for spoofing the receiver, and then assign the attack type in \verb+ModelTypes+ as intent ``spoofing". In this case, the sender will let the receiver do something that the sender cannot do without the necessary permission for it. Therefore, the receiver's privileges will be misused unexpectedly.

\textbf{Case 4} (line 16-17): When the receiver starts a private component with the permissions for the data from the sender, we think the receiver is spoofed and thus add the model and the attack type to \verb+ModelTypes+. Under this circumstance, the sender calls a private component via the other exposed components and will not trigger any illegal behavior.

\textbf{Case 5} (line 18-19): When the sender lacks the permissions that the receiver needs for calling a sensitive method, meanwhile the receiver also lacks the permission for generating data from the sender, we consider that they are colluding and thus add the type intent ``\emph{collusion}" attack to \verb+ModelTypes+, indicating that they are escalating the privilege from each other as well as complementing permissions for each other through the inter-component communication process, which is a clear case of intent collusion attack.

\subsection{The Complexity}
\label{complexity}

The complexity of SIAT depends on the number of apps and components at runtime. Since the complexity of \emph{Monitor} mainly relies on the actual lifetime of the app, thus here we focus on analyzing the complexity of \emph{Analyzer}. Assume all feasible ICC models between $n$ apps which per app contains $m$ components need to be analyzed in the SIAT. For every component, there are $(nm-1)$ components to communicate with. In practical, if only the vulnerable paths between two different apps could incur malicious attacks, there are $nm \choose 1$$\times$$(n-1)m \choose 1$ models in this situation. Therefore, the computation complexity of building the ICC Models is $O(n^{2}m^{2})$, the complexity of identifying the threat model is $O(n^{2}m^{3})$.

\begin{table*}[h]
\small
\caption{Overview of comparisons of accuracy between DIALDroid, Amandroid, XManDroid and SIAT.}
\label{table:accuracy}
\begin{center}
\addvbuffer[-20pt 0pt]{
\resizebox{\textwidth}{!}{%
\begin{tabular}{l|c|c|c|c|c|c|c|c|c|c|c|c|c|c}
\hline
\multicolumn{1}{c|}{\multirow{2}{*}{\textbf{DataSet}}} & \multicolumn{2}{c|}{\textbf{Number}} & \multicolumn{4}{c|}{\textbf{Malicious ICC Paths($\sum$$\checkmark$/$\sum$$\otimes$)}} & \multicolumn{4}{c|}{\textbf{Precision($p=\sum\checkmark/(\sum\checkmark+\sum\otimes)$)/Recall($r=\sum\checkmark$/number of ICC)}} & \multicolumn{4}{c}{\textbf{F-Measure}($2pr/(p+r)$)} \\ \cline{2-15}
\multicolumn{1}{c|}{} & Apps & ICC & DIALDroid & Amandroid & XManDroid & SIAT & DIALDroid & Amandroid & XManDroid & SIAT & DIALDroid & Amandroid & XManDroid & SIAT \\ \hline\hline

\textbf{DroidBench3.0} & 11 & 9 & 7/2 & 0/0 & 9/0 & 9/0 & 0.78/0.78 & 0/0 & 1.00/1.00 & 1.00/1.00 & 0.75 & 0.00 & 0.94 & 1.00 \\ \hline
\textbf{IccTA} & 6 & 3 & 3/0 & 3/0 & 3/0 & 3/0 & 1.00/1.00 & 1.00/1.00 & 1.00/1.00 & 1.00/1.00 & 1.00 & 1.00 & 1.00 & 1.00 \\ \hline
\textbf{Our Developments} & 40 & 26 & 10/2 & 5/0 & 19/2 & 26/0 & 0.83/0.38 & 1.00/0.19 & 0.90/0.73 & 1.00/1.00 & 0.52 & 0.32 & 0.81 & 1.00 \\ \hline
\textbf{Real-World} & 75 & 8 & 5/10 & 2/4 & 6/8 & 7/0 & 0.33/0.63 & 0.33/0.25 & 0.43/0.75 & 1.00/0.88 & 0.43 & 0.28 & 0.55 & 0.93 \\ \hline
\textbf{Total} & \cellcolor[rgb]{.9,.8,.9}132 & \cellcolor[rgb]{.9,.8,.9}46 & \cellcolor[rgb]{.9,.8,.9}25/14 & \cellcolor[rgb]{.9,.8,.9}10/4 & \cellcolor[rgb]{.9,.8,.9}37/10 & \cellcolor[rgb]{.9,.7,.9}45/0 & \cellcolor[rgb]{.9,.8,.9}0.64/0.54 & \cellcolor[rgb]{.9,.8,.9}0.71/0.21 & \cellcolor[rgb]{.9,.8,.9}0.78/0.80 & \cellcolor[rgb]{.9,.7,.9}1.00/0.98 & \cellcolor[rgb]{.9,.8,.9}0.59 & \cellcolor[rgb]{.9,.8,.9}0.32 & \cellcolor[rgb]{.9,.8,.9}0.79 & \cellcolor[rgb]{.9,.7,.9}0.99 \\ \hline
\end{tabular}%
}
}
\end{center}
\footnotesize{ The ICC path in Table \ref{table:accuracy} and  \ref{table:bench_details} denotes the malicious ICC path incurring attack behaviors. $\checkmark$=True Positive, $\otimes$=False Positive, $\odot$=False Negative.}
\end{table*}

\begin{table*}[h]
\small
\caption{The partial results of ICC paths detection in DroidBench3.0, IccTA and Our Developments. The attack behaviors of Real-World are listed in the Table \ref{table:realworld}.}
\label{table:bench_details}
\begin{center}
\addvbuffer[-20pt -15pt]{
\resizebox{\textwidth}{!}{%
\begin{tabular}{c|l|l|c|c|c|c|c}
\hline
\textbf{Dataset} & \multicolumn{1}{c|}{\textbf{Source}} & \multicolumn{1}{c|}{\textbf{Destination}} & \multicolumn{1}{l|}{\textbf{Num. of ICC Paths}} & \multicolumn{1}{l|}{\textbf{DIALDroid}} & \multicolumn{1}{l|}{\textbf{Amandroid}} & \textbf{XManDroid} & \multicolumn{1}{l}{\textbf{SIAT(Ours)}} \\ \hline \hline
\multirow{9}{*}{\textbf{DroidBench3.0}} & SendSMS & Echoer & 1 & \cellcolor[rgb]{.9,.7,.9}$\checkmark$ $\otimes$ & $\odot$ & $\checkmark$ & \cellcolor[rgb]{.9,.8,.9}$\checkmark$ \\ \cline{2-8}
 & StartActivityForResult1 & Echoer & 1 & \cellcolor[rgb]{.9,.7,.9}$\checkmark$ $\otimes$ & $\odot$ & $\checkmark$ & \cellcolor[rgb]{.9,.8,.9}$\checkmark$ \\ \cline{2-8}
 & DeviceId\_Broadcast1 & Collector & 1 & \cellcolor[rgb]{.9,.8,.9}$\checkmark$ & $\odot$ & $\checkmark$ & \cellcolor[rgb]{.9,.8,.9}$\checkmark$ \\ \cline{2-8}
 & DeviceId\_ContentProvider1 & Collector & 1 & \cellcolor[rgb]{.9,.8,.9}$\checkmark$ & $\odot$ & $\checkmark$ & \cellcolor[rgb]{.9,.8,.9}$\checkmark$ \\ \cline{2-8}
 & DeviceId\_OrderedIntent1 & Collector & 1 & \cellcolor[rgb]{.9,.8,.9}$\checkmark$ & $\odot$ & $\checkmark$ & \cellcolor[rgb]{.9,.8,.9}$\checkmark$ \\ \cline{2-8}
 & DeviceId\_Service1 & Collector & 1 & \cellcolor[rgb]{.9,.7,.9}$\odot$ & $\odot$ & $\checkmark$ & \cellcolor[rgb]{.9,.8,.9}$\checkmark$ \\ \cline{2-8}
 & Location1 & Collector & 1 & \cellcolor[rgb]{.9,.8,.9}$\checkmark$ & $\odot$ & $\checkmark$ & \cellcolor[rgb]{.9,.8,.9}$\checkmark$ \\ \cline{2-8}
 & Location\_Broadcast1 & Collector & 1 & \cellcolor[rgb]{.9,.8,.9}$\checkmark$ & $\odot$ & $\checkmark$ & \cellcolor[rgb]{.9,.8,.9}$\checkmark$ \\ \cline{2-8}
 & Location\_Service1 & Collector & 1 & \cellcolor[rgb]{.9,.7,.9}$\odot$ & $\odot$ & $\checkmark$ & \cellcolor[rgb]{.9,.8,.9}$\checkmark$ \\ \hline \hline
 \multirow{3}{*}{\textbf{IccTA}} & startActivity1\_source & startActivity1\_sink & 1 & \cellcolor[rgb]{.9,.8,.9}$\checkmark$ & $\checkmark$ & $\checkmark$ & \cellcolor[rgb]{.9,.8,.9}$\checkmark$ \\ \cline{2-8}
 & startService1\_source & startService1\_sink & 1 & \cellcolor[rgb]{.9,.8,.9}$\checkmark$ & $\checkmark$ & $\checkmark$ & \cellcolor[rgb]{.9,.8,.9}$\checkmark$ \\ \cline{2-8}
 & startbroadcast1\_source & startbroadcast1\_sink & 1 & \cellcolor[rgb]{.9,.8,.9}$\checkmark$ & $\checkmark$ & $\checkmark$ & \cellcolor[rgb]{.9,.8,.9}$\checkmark$ \\ \hline \hline
 \multirow{10}{*}{\textbf{Our Developments}} & Sender0 & ReceiverTest0 & 1 & \cellcolor[rgb]{.9,.8,.9}$\checkmark$ & $\checkmark$ & $\checkmark$ & \cellcolor[rgb]{.9,.8,.9}$\checkmark$ \\ \cline{2-8}
 & Sender0 & ReceiverTest1 & 1 & \cellcolor[rgb]{.9,.8,.9}$\checkmark$ & $\checkmark$ & $\checkmark$ & \cellcolor[rgb]{.9,.8,.9}$\checkmark$ \\ \cline{2-8}
 & Sender0 & ReceiverTest2 & 1 & \cellcolor[rgb]{.9,.8,.9}$\checkmark$ & $\checkmark$ & $\checkmark$ & \cellcolor[rgb]{.9,.8,.9}$\checkmark$
 \\ \cline{2-8}
 & Sender0 & Receiver-shareference & 1 & \cellcolor[rgb]{.9,.7,.9}$\odot$ & \cellcolor[rgb]{.9,.7,.9}$\odot$ & \cellcolor[rgb]{.9,.7,.9}$\odot$ & \cellcolor[rgb]{.9,.7,.9}$\checkmark$
 \\ \cline{2-8}
 & Sender0 & Receiver-application & 1 & \cellcolor[rgb]{.9,.7,.9}$\odot$ & \cellcolor[rgb]{.9,.7,.9}$\odot$ & \cellcolor[rgb]{.9,.7,.9}$\odot$ & \cellcolor[rgb]{.9,.7,.9}$\checkmark$
 \\ \cline{2-8}
 & Sender1 & ReceiverTest0 & 1 & \cellcolor[rgb]{.9,.8,.9}$\odot$ & $\odot$ & $\checkmark$ & \cellcolor[rgb]{.9,.8,.9}$\checkmark$ \\ \cline{2-8}
 & Sender1 & ReceiverTest1 & 1 & \cellcolor[rgb]{.9,.8,.9}$\odot$ & $\odot$ & $\checkmark$ & \cellcolor[rgb]{.9,.8,.9}$\checkmark$ \\ \cline{2-8}
 & Sender1 & ReceiverTest2 & 1 & \cellcolor[rgb]{.9,.8,.9}$\odot$ & $\odot$ & $\checkmark$  & \cellcolor[rgb]{.9,.8,.9}$\checkmark$
 \\ \cline{2-8}
 & Sender1 & Receiver-shareference & 1 & \cellcolor[rgb]{.9,.7,.9}$\odot$ & \cellcolor[rgb]{.9,.7,.9}$\odot$ & \cellcolor[rgb]{.9,.7,.9}$\odot$ & \cellcolor[rgb]{.9,.7,.9}$\checkmark$
 \\ \cline{2-8}
 & Sender1 & Receiver-application & 1 & \cellcolor[rgb]{.9,.7,.9}$\odot$ & \cellcolor[rgb]{.9,.7,.9}$\odot$ & \cellcolor[rgb]{.9,.7,.9}$\odot$ & \cellcolor[rgb]{.9,.7,.9}$\checkmark$
 \\ \hline
\end{tabular}%
}}
\end{center}
\end{table*}


\section{Evaluations}
\label{evaluation}

This section presents the experimental evaluation results of SIAT based on the four datasets below:

\begin{itemize}
\item DroidBench3.0 \cite{benchdroid3.0}, which is an app collection for benchmarking ICC-based sensitive data leaks and consists of many types of ICC-related attacks.
\item IccTA \cite{droidbench-iccta}, which has three sets of apps for testing the inter-app collusion issues, and was released by EC SPRIDE Secure Software Engineering Group.
\item Our Developments, which is similar to the DroidBench3.0 and consists of more than forty self-developed apps that only have simple threat models and functions for comprehensive testing, and twenty-six ICC processes covering at least three components with various sensitive APIs. Concerning the efficiency and accuracy, the intent call entries are consistent with the app entries to simplify the call graph, and each app-pair ICC is independent of each other.
\item Real-World, which contains about 2100 real-world apps downloaded from the Google Play market\cite{googleplay}.
\end{itemize}

Our evaluation addresses the following three questions:

\begin{itemize}
    \item RQ1. What is the accuracy of SIAT compared to state-of-the-art approaches?
    \item RQ2. How well does SIAT perform in practice? What could SIAT find in real-world applications?
    \item RQ3. What about the individual performance of the \emph{Monitor} and the \emph{Analyzer}?
\end{itemize}


\subsection{Results for RQ1 (Accuracy)}
We compare the SIAT with the state-of-the-art approaches introduced in Section \ref{related work} (i.e., well-known static approaches DIALDroid and AmanDroid, and runtime technology XManDroid) for ICC vulnerability detection.

\begin{figure}[tb]
\centerline{\includegraphics[scale=0.45]{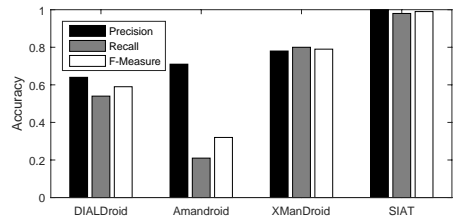}}
\caption{Comparisons of three accuracy metrics.}
\label{accuracy in total}
\end{figure}

Figure \ref{accuracy in total} and Table \ref{table:accuracy} provide an overview of the comparisons of accuracy. The details of ICC paths detected are listed in Tables \ref{table:bench_details} and \ref{table:realworld}. In particular, to ensure that the inter-app attacks could be launched in Real-World, as we will explain in Section \ref{subsec:RQ2}, we have analyzed thousands of suspicious apps to eliminate most of them. For instance, we firstly manually inspected the codes of every application to investigate if each identified ICC path was indeed vulnerable. After injecting some codes into the Android system, we can further identify via the system debug outputs if the malicious activities have been launched to exploit the vulnerable application successfully. Although we have identified more than twenty malicious ICC paths in Real-World, here we randomly choose seventy-five apps from Real-World which only cover eight ICC paths, as shown in Table \ref{table:accuracy} and \ref{table:realworld}.

\vspace{-0.1in}

\subsubsection{Accuracy Comparisons Overview}
As depicted in Figure \ref{accuracy in total} and Table \ref{table:accuracy}, we employ three performance indicators to evaluate the accuracy: \textbf{Precision}, denoted by $p$, which is the fraction of related instances among the retrieved instances; \textbf{Recall}, denoted by $r$, which is the fraction of the total amount of relevant instances that are actually retrieved; \textbf{F-measure} \cite{f-measure}, which computes the comprehensive score $\frac {2\times p\times r}{p+r}$. Figure \ref{accuracy in total} depicts the value of three indicators in total. It is worth noting that the \emph{ICC path} in this section denotes the malicious ICC path incurring attack behaviors.

As we can see, in Figure \ref{accuracy in total}, the DIALDroid merely obtains 0.64 precision and 0.54 recall in total. The DAILDroid performs static taint analysis to identify attributes of the intent and the intent-filter, to trace the data flow associated with the intent. Then it uses SQL stored procedures and queries to calculated sensitive channels in the database according to the matching rules between the intent and the intent-filter. However, the DAILDroid cannot accurately tell whether the data in the intent meets the requirements of the receiving component. When the data format doesn't meet the requirements in the program, the sensitive method will not be executed. But the DAILDroid does not consider it and determines that the sensitive method must be executed if the intent attributes do match. In addition, DIALDroid treats the cases that sensitive data arrives in other applications via intent as a privacy breach, which improves the overall coverage while introducing false positives. Similarly, the AmanDroid achieves 0.71 precision and 0.21 recall in that it cannot analyze the data flow when dealing with the complex ICC paths.

As shown in Figure \ref{accuracy in total} and Table \ref{table:accuracy}, the XManDroid obtains 0.73 recall on Our Developments due to the seven ICC paths suffering from intent spoofing, which the XManDroid cannot identify. Consequently, the XManDroid only achieves a precision of 0.78 and a recall of 0.80 in total.
The XManDroid enables users to predefine a list of ICC restriction policies and automatically block ICCs that match with any policy. These policies are based on the permissions in the sender and the data in intent. Thanks to its permission identification mechanism, which will not intercept the deliver of intent only if the permissions in the receiver match the ones in the sender, the XManDroid performs well both on IccTA and DroidBench3.0, as shown in Table \ref{table:accuracy}. However, when the sender sends out the sensitive data with the permission that the receiver doesn't have, the XManDroid prohibits this ICC directly without considering whether the receiver uses the data later. \textit{This case is a common problem in many runtime protection approaches, which raises a high false alarm rate}. For example, the experimental results on Our Developments shows that even the receiver does not extract any sensitive data from intent, the XManDroid still thinks there is a malicious behavior without identifying the receiver's behaviors, which makes the XManDroid detect two false positives $\otimes$. In contrast, the SIAT traces both of the data flows in the senders and receivers, then analyzes the whole transmission process that enables SIAT to generate less false negatives than the XManDroid.

Compared to the existing approaches, as depicted in Figure \ref{accuracy in total} and Table \ref{table:accuracy}, the proposed SIAT can achieve about 25\%$\sim$200\% accuracy improvements with 1.0 precision and 0.98 recall in total, even though it is unable to trace a few output methods due to the limits of the built-in TaintDroid. \textit{There are two reasons why SIAT performs much better: firstly, unlike the DAILDroid and the XManDroid, SIAT traces the data flow in the receiver at runtime through capturing and verifying the data in sensitive method, which makes SIAT acquire more precise data flows; secondly, DIALDroid, AmanDroid, and XManDroid do not detect intent spoofing, which is one of the major reasons why their precision is lower than ours. Therefore, SIAT is able to achieve the analysis results which are more closer to the real situations happened between apps than DAILDroid in the above scenario.}

\subsubsection{Details of ICC Path Detection}
Table \ref{table:bench_details} shows details of malicious ICC paths of DroidBench 3.0 and IccTA, and only ten malicious ICC paths of Our Developments due to the paper limits. The ICC paths in Real-World are given in Section \ref{subsec:RQ2}. It is worth noting that the original three ICC paths in IccTA are innocent since the receivers can get the device ID from intent by itself with the related permissions. To make the ICC paths illegal, we delete the permissions for device ID in the three receivers.

The results in DroidBench 3.0 for DIALDroid are much better than AmanDroid; nevertheless, there still are two deficiencies: the first one is that the DIALDroid cannot identify the malicious ICC path when the type of the component is \verb+Service+; obviously, there are two cases for the two source apps named \verb+DeviceId_Service1+ and \verb+Location_Service1+; the second one is that the DIALDroid simply considers all possible branches to be executed when facing many branches in source codes, e.g., for destination app named \verb+Echoer+, the two branches in codes make the DIALDroid detect two false malicious ICC paths.

Furthermore, as mentioned before, the DIALDroid ignores the receiver's real requirements of the intent data formats, leading to extra false positives. For instance, in Real-World dataset, for app \verb+vbox7handler+, it will exit immediately if the data in the intent does not have \verb+vbox7 $\backslash$.com $\backslash/$play+. However, DIALDroid still constructs the vulnerable ICC path between the sender and the \verb+vbox7handler+. For another app named \verb+UrlToPdf+, after receiving the intent, it will output the data to logs only if it identifies there is a key-value pair with the key \verb+android.intent.extra.TEXT+ in the vector \verb+EXTRA+ of the data in the intent. Nevertheless, the DIALDroid still considers that the log should be triggered since it doesn't care if the receiver validates the data or not. While SIAT can distinguish whether or not the log function should be triggered.

On the other hand, all four approaches can achieve good detection results on IccTA, regarding the simple ICC paths in 3-pair apps. Since the AmanDroid cannot analyzer the data flows when facing the complex ICC paths, it cannot detect any malicious ICC path in DroidBench 3.0.

\begin{figure}[tb]
\centerline{\includegraphics[scale=0.55]{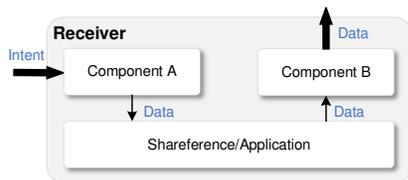}}
\caption{Two cases of bypassing in the receiver.}
\label{two case}
\end{figure}

\subsubsection{Two Cases of Bypassing}
The bypassing is similar to the malware collusion in a way, i.e., two components try to work cooperatively so that each component only performs part of the behavior to bypass the detection. Nevertheless, the main difference is that the two components come from the same application in the above cases of bypassing. Based on extensive experiments and in depth analysis, as depicted in Figure \ref{two case}, in the receiver, we discover the following two undisclosed cases of malicious bypassing which can invalidate the existing approaches by taking advantage of especial intermediate methods/objects:

The first case is that, as shown in Figure \ref{two case}, in the receiver, if the \emph{Component A} stores the sensitive data into the \emph{Shareference} in the form of a key-value pair after receiving the intent. Subsequently, the \emph{Component B} can extract the data from the \emph{Shareference} object and further output the data to the outside of the device. Listing \ref{sf-a} and \ref{sf-b} present example codes to showcase the bypassing in this case.
\begin{lstlisting}[caption={The Component A puts the sensitive data into Shareference.},label=sf-a,float,breaklines,columns=flexible]
public class Component_A extends AppCompatActivity { protected void onCreate(Bundle savedInstanceState) { ...
    Intent receivedIntent=getIntent();
    receivedIntent.setClass(this, MyService.class);
    Editor editor=getSharedPreferences("settings", 0).edit();
    String id=receivedIntent.getStringExtra("android.intent.extra.TEXT");
    if (id != null) {
      editor.putString("deviceId", id);
      editor.commit();
    }
    startService(receivedIntent);
  }
}
\end{lstlisting}

Similarly, in the second case, \emph{component A} assigns the data extracted from the intent to the variable of the \emph{Application} object after receiving the intent. It is worth noting that there is a unique Application object per Android app at runtime so that each component can find the same one. Afterward, another \emph{component B} can extract the data from the intent by searching the variable in the \emph{Application} object, and then calls the APIs for sending the SMS or writing in a file to output the data to the outside of the device as aforementioned.

\begin{lstlisting}[caption={The Component B gets the sensitive data from Shareference and then outputs them.},label=sf-b,float,breaklines,columns=flexible]
public class Component_B extends Service {
  public int onStartCommand(Intent intent, int flags, int startId) {
    Log.v("leakData", getSharedPreferences("settings", 0).getString("deviceId", "default"));
    return super.onStartCommand(intent, flags, startId);
  }
}
\end{lstlisting}

\begin{figure*}[!t]
\centering
\includegraphics[scale=0.7]{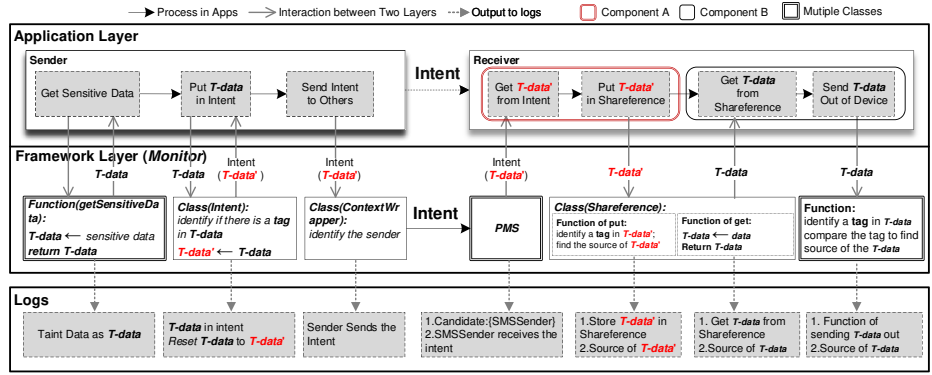}
\caption{The \emph{Monitor}'s workflow on examples of Listings \ref{sf-a} and \ref{sf-b}.}
\label{monitor-sample}
\end{figure*}

We have successfully realized the above two bypassing cases in Our Developments with destination named \verb+Receiver-shareference+ and \verb+Receiver-application+, as shown in Table \ref{table:bench_details}. For the first case, both DAILDroid and AmanDroid only can notice that the \emph{Component A} has stored the sensitive data in the \emph{Shareference}, but cannot detect that \emph{Component B} has obtained the sensitive data from the \emph{Shareference}, and its following malicious behaviors. For the second case, we employ DIALDroid and AmanDroid to try to detect the app-pair in several detection rounds. However, the results obtained by the two approaches show that there is no leak of sensitive data from the intent in the receiver. Consequently, malicious apps can easily bypass the detection of DIALDroid and AmanDroid in this case. Also, the XManDroid cannot detect the two cases due to its omitting of the actual behavior of the receiver, which incurs the false negatives $\odot$.

\begin{algorithm}[h]
\label{bypass}
\caption{Identification of original source of tainted data for bypassing}
\renewcommand{\algorithmicrequire}{\textbf{Input:}}	\renewcommand{\algorithmicensure}{\textbf{Output:}}
\begin{algorithmic}[1]
\REQUIRE $m \Leftarrow$ current model needs to find original source of tainted data
\ENSURE $m \Leftarrow$ improved model contains the original source of tainted data;
\STATE Let $models$ be the models generated before
\STATE Let $model$ be a model in $models$.
\STATE Let $sender$ be the sender object in a model.
\STATE Let $intent$ be the intent object in a model.
\STATE Let $rcver$ be the receiver object in a model.
\STATE Let $tdata$ be the tainted data in a model.
\STATE Let $source[tdata]$ be the source method of $tdata$.

\FOR{each $tdata \in m.rcver.tdatas$}
\IF{($tdata \in m.intent.tdatas)$}
\STATE {$m.source[tdata]$$\leftarrow$$model.sender$}
\ENDIF

\FOR{each $model \in models$}
\IF{($model.rcver.component$=$m.sender.component)\wedge(tdata \in model.intent.tdatas$)}
\STATE {$m.source[tdata]$$\leftarrow$$model.source[tdata]$}
\STATE $continue$
\ENDIF
\ENDFOR

\FOR{each $model \in models$}
\IF{$tdata \in model.intent.tdatas$}
\STATE {$m.source[tdata]$$\leftarrow$$model.source[tdata]$}
\ENDIF
\ENDFOR
\ENDFOR
\end{algorithmic}
\end{algorithm}

Based on the workflow for monitoring \textit{Shareference} in Figure \ref{monitor-sample} and the identification algorithm of the original sources of tainted data for the bypassing in Algorithm 2, we showcase the solution of bypassing as follows:

Depending on a systemwide real-time tracing of the tainted data for \textit{Shareference}, as shown in Figure \ref{monitor-sample}, the \emph{Monitor} firstly taints the original sensitive data with a tag and then retaints the data as \emph{T-data}$'$ (\emph{T-data}$'$ $\leftarrow$ \emph{T-data}) when storing it in an intent. After being delivered to the receiver with intent, whatever it has experienced in the receiver, only if the \emph{T-data}$'$ is utilized as the parameter of some sensitive function which is used to store data and then output them out of the device, our \emph{Monitor} will identify the \emph{T-data}$'$ and figure out its original source by comparing the taint tag and the predefined source code. In this way, as depicted in Figure \ref{monitor-sample}, when the corresponding component puts or gets the \emph{T-data}$'$ via the \verb+Class(Shareference)+, and further sends the \emph{T-data}$'$ out of the device, the functions that use the \emph{T-data}$'$ will be monitored by the \emph{Monitor}. The \emph{Monitor} is able to  discover the \emph{T-data}$'$ as the tainted parameter immediately and then search the tag in it to find out the original sender of it. Afterward, based on the improved models built by the Algorithm 2, which is adopted to identify the original sources of the tainted data with taint logs given the intermediate methods/objects, the \emph{Analyzer} is capable of determining whether the \emph{T-data}$'$ in the functions above are the \emph{T-data}$'$ in the sender's intent according to the information provided by the \emph{Monitor}. It is worth noting that, different from the \textit{Application}, we implement the Algorithm 2 for the \textit{Shareference} through a taint value matching mechanism to trace the data flow based on the particular key-value pairs putting/getting entries in logs. To improve the original sources of the tainted data in the models, Algorithm 2 matches the tainted data in intent with the data in the receiver iteratively.

Therefore, both the functions utilized to store the \emph{T-data}$'$ in \emph{Application} or \emph{Shareference} in \emph{Component A}, and the functions utilized to get \emph{T-data}$'$ from \emph{Application} or \emph{Shareference} in \emph{Component B}, cannot eliminate the taint tag in \emph{T-data}$'$ so that the malicious behaviors never could bypass our approach to send sensitive data out of the device.

\begin{table*}[h]
\begin{center}
\caption{Analysis results on real-world apps.}
\label{table:realworld}
\resizebox{\textwidth}{!}{%
\addvbuffer[-15pt 0pt]{
\begin{tabular}{c|c|c|c|c|c|c|c|c}
\hline
\multicolumn{4}{c|}{\textbf{Sender}} & \multicolumn{4}{c|}{\textbf{Receiver}} & \multirow{2}{*}{\textbf{Type}} \\ \cline{1-8}
\textbf{Name} & \textbf{Version} & \textbf{Component} & \textbf{Sensitive Data} & \textbf{Name} & \textbf{Version} & \textbf{Component} & \textbf{Sensitive Method} &  \\ \hline\hline
TopGoodNightImages & 10 & StartActivity & string-extra & TraductoresScout & 33 & MainActivity & Log & \multicolumn{1}{l}{\cellcolor[rgb]{.9,.8,.9}\textit{\textbf{hijacking}}} \\ \hline
PEC2012 & 24 & MoreTabActivity & string-extra & Prizmshare & 2 & MainActivity & write & \multicolumn{1}{l}{\cellcolor[rgb]{.9,.8,.9}\textit{\textbf{spoofing}}} \\ \hline
SimCardManager & 20400 & Main & deviceId & Notepad & 50 & NoteEditActivity & fileOutputStream & \multicolumn{1}{l}{\cellcolor[rgb]{.9,.8,.9}\textit{\textbf{collusion}}} \\ \hline
fotoalbumgpslite & 8 & ImageActivity & location & silentcamera & 13 & CameraActivity & --- & \cellcolor[rgb]{.9,.8,.9}--- \\ \hline
lowlevel.sendapp & 11 & Application & --- & urlripper & 72 & ProcessIntent & --- & \cellcolor[rgb]{.9,.8,.9}--- \\ \hline
\end{tabular}%
}
}
\end{center}
\footnotesize{The string-extra here denotes the string \verb+extra+ field in intent. `---' denotes the false positive case.}
\end{table*}

\subsection{Results for RQ2 (How SIAT performs on Real-World Apps)}
\label{subsec:RQ2}

We run Real-World applications by adopting the automated testing script to trigger the applications' behaviors in the system for detection purposes. For each app, we have to write the corresponding scripts to run the test with \emph{monkeyrunner} \cite{monkeyrunner}, which is a popular Android tool for running test suites. It is a time-consuming task to handle a large number of apps in this way. To save the testing time and effort for the apps without ICCs, we first exploit the static analysis approaches to find out the apps that holding the ICC paths, and then analyze the other apps one by one and write the corresponding scripts for them; finally, we run the scripts and applications simultaneously in our system, so as to trigger the application behaviors for analysis. After analyzing about 2100 suspicious apps of Real-World, we have excluded most of them. We find that there are ICCs in 163 application pairs without suspicious behavior. For all pairs of applications that have ICCs, there are no sensitive data in the ICCs of the 121 applications. On the other hand, there are intent hijacking attacks in the ICCs of sixteen application pairs, intent spoofing attacks in the ICCs of six application pairs, and  malware collusions in the ICCs of four application pairs. Table \ref{table:realworld} illustrates several unrevealed instances of the threats mentioned above as well as false positives identified by SIAT.

\textbf{Intent hijacking}. The \verb+TopGoodNightImages+ (10,000+ downloads) sends out an intent whose \verb+extra+ field stores the non-sensitive data. While the \verb+TraductoresScout+ (50,000+ downloads) receives the intent with required attributes and then lets the data leak out from intent into a log. Thus the data from the \verb+TopGoodNightImages+ can be sent out of the \verb+TraductoresScout+, and other sensitive data will make it even worse.

\textbf{Intent spoofing}. After the \verb+Prizmshare+ receiving the intent from the \verb+PEC2012+, it writes the data extracted from the intent into a file. Even though the \verb+PEC2012+ can't write data to a file by itself due to the lack of the write permission, the receiver \verb+Prizmshare+ can escalates the privilege for it.

\textbf{Intent collusion attack}. The \verb+SimCardManager+ (10,000+ downloads) sends out an intent with device ID to the receiver \verb+Notepad+ (1,000,000+ downloads), then the \verb+Notepad+ receives the intent and stores the extracted device ID into a file. Since the \verb+SimCardManager+ doesn't have the permissions to store a file and the \verb+Notepad+ neither has the permission to get the devices ID. Therefore it's utterly suspicious that they are complementing the permissions for each other.

\textbf{False positive cases}. Besides, there are two typical cases of false positives in existing approaches, which are addressed by SIAT in Table \ref{table:realworld}. The \verb+Fotoalbumgpslite+ sends out an intent whose data has the device's locations. While for the \verb+silentcamera+ (5,000,000+ downloads), the location information in intent is never traced in storage methods (e.g. \verb+Shareference.putString+) or other sensitive methods. Therefore, the sensitive data will not be leaked out to the receiver. The \verb+Lowlevel+ sends out an URL to the \verb+urlripper+, and then the \verb+urlripper+ is able to access the Internet address. Hence, in SIAT, it is legal for the sender to utilize the function of other applications owing to its Internet permission.

\subsection{Results for RQ3 (Run-time Performance)}

\begin{figure*}
\subfigure[Real-World.]{
\label{performance of monitor:realapp}
\begin{minipage}[b]{0.33\textwidth}
\centering
\includegraphics[scale=0.4]{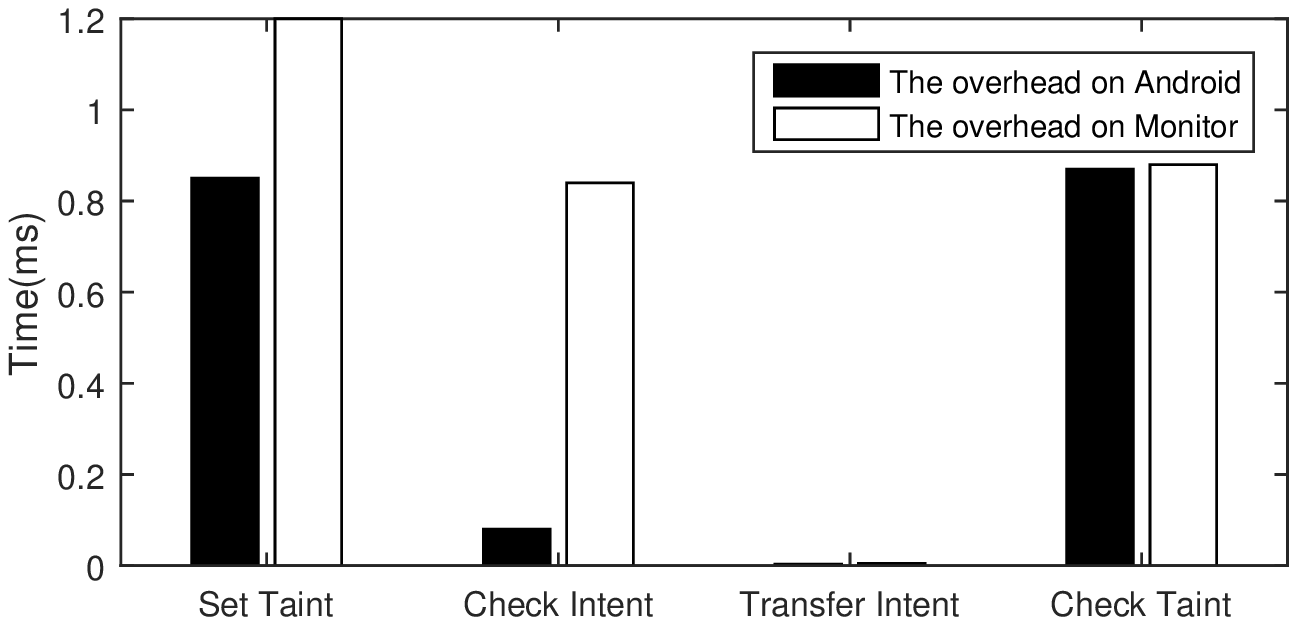} \\
\end{minipage}}%
\subfigure[Our Developments.]{
\label{performance of monitor:ours} 
\begin{minipage}[b]{0.33\textwidth}
\centering
\includegraphics[scale=0.4]{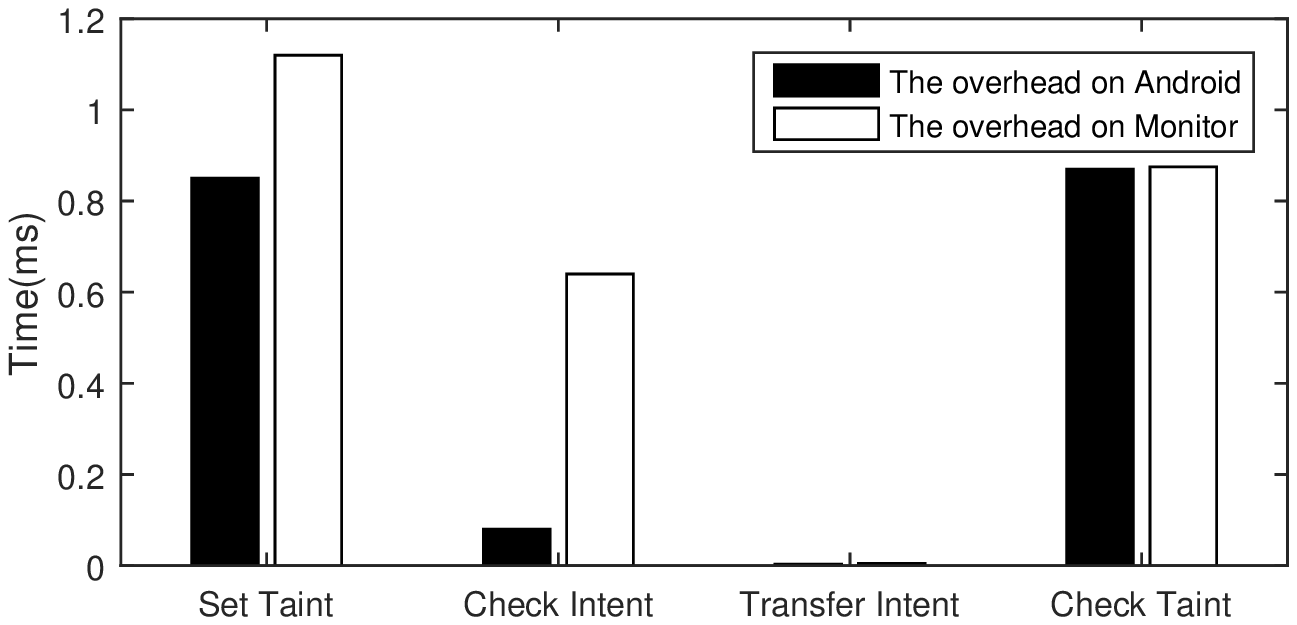} \\
\end{minipage}}
\subfigure[DroidBench3.0+IccTA.]{
\label{performance of monitor:bench} 
\begin{minipage}[b]{0.33\textwidth}
\centering
\includegraphics[scale=0.4]{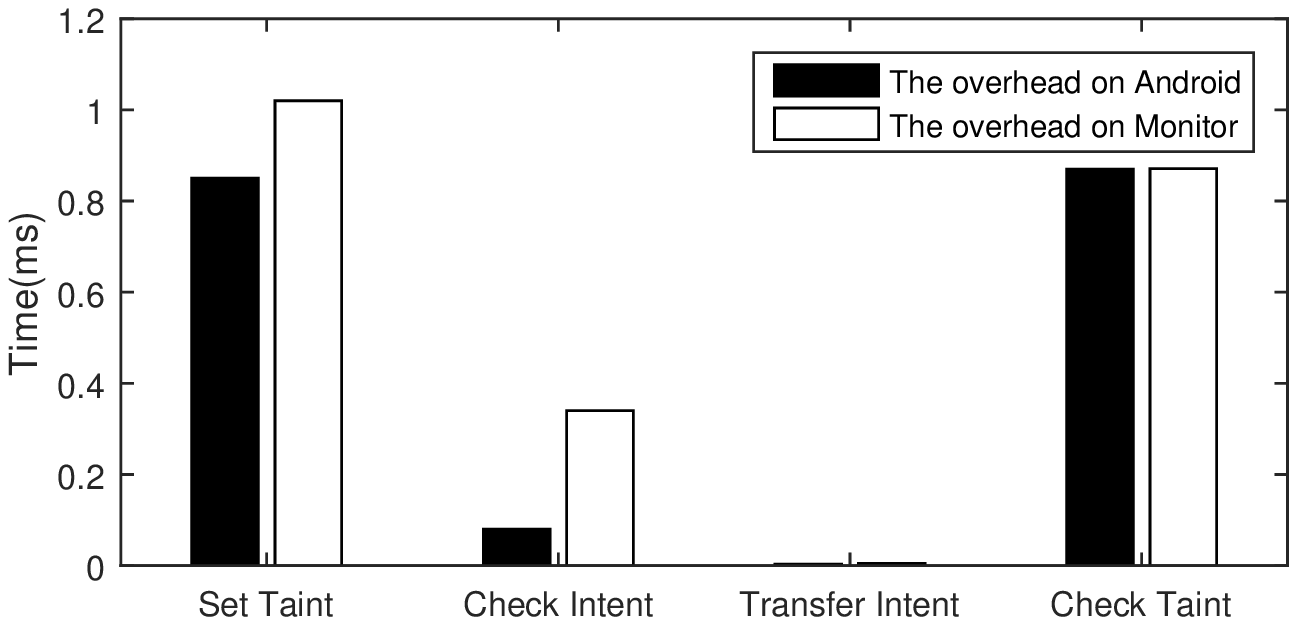} \\
\end{minipage}}
\caption{The comparison of app runtime in various parts between \emph{Monitor} and Android on different datasets.}
\label{monitor runtime cost} 
\end{figure*}

We now evaluate the runtime performance of SIAT with three datasets, i.e., DroidBench3.0+IccTA, Our Developments, and Real-World. Since the \emph{Monitor} and the \emph{Analyzer} are two independent components on the workflow of SIAT. Notably, the actual runtime of \emph{Monitor} is related to the lifetime of app, hence we evaluate their runtime overhead below separately before summing them.

\subsubsection{Monitor Performance}
Since our \emph{Monitor} is a modified version of the Android system at framework layer, we compute the app runtime cost on our \emph{Monitor} and the native Android operating system, respectively. Particularly, we have randomly selected dozens of app-pairs that can launch various ICC attacks from the three datasets mentioned above, respectively. Also, to achieve the accurate runtime interval, we have inserted related time-stamped recording codes in a variety of crucial APIs in the \emph{Monitor} and apps. It is worth noting that the runtime cost of each part on Android in Figure \ref{monitor runtime cost} represents the recorded app execution time at the same APIs after we running apps without the \emph{Monitor}. We then run apps under the \emph{Monitor} and the Android native operating system, respectively. We divide the \emph{Monitor} module into four parts according to the five steps in Section \ref{subsec:arch-detail} to calculate and compare the time cost separately. We run apps on two systems about twenty times, respectively, and adopt the average value as the final execution time.

Figure \ref{monitor runtime cost} shows the evaluation results for each part. As we can see in Figure \ref{performance of monitor:realapp}, the time overhead of Real-World is apparently longer than the others in that the real world apps maintain more complex and complete functionalities that lead to more ICC paths. Specifically, both in the transfer intent and check taint processes, the functions for \emph{Monitor} almost doesn't incur any runtime overhead compared with original Android. The set taint process leads to about 0.3ms overhead due to the import of TaintDroid functions. The major runtime cost is incurred by the check intent process in \emph{Monitor}, which exploits the reflective calls to figure out the component who sends the intent. Nevertheless, the overhead is less than 1ms and thus is negligible for the end-users.

\begin{figure*}
\subfigure[Varying Number of App-pairs.]{
\label{total time of analysis:varying number}
\begin{minipage}[b]{0.45\textwidth}
\centering
\includegraphics[scale=0.4]{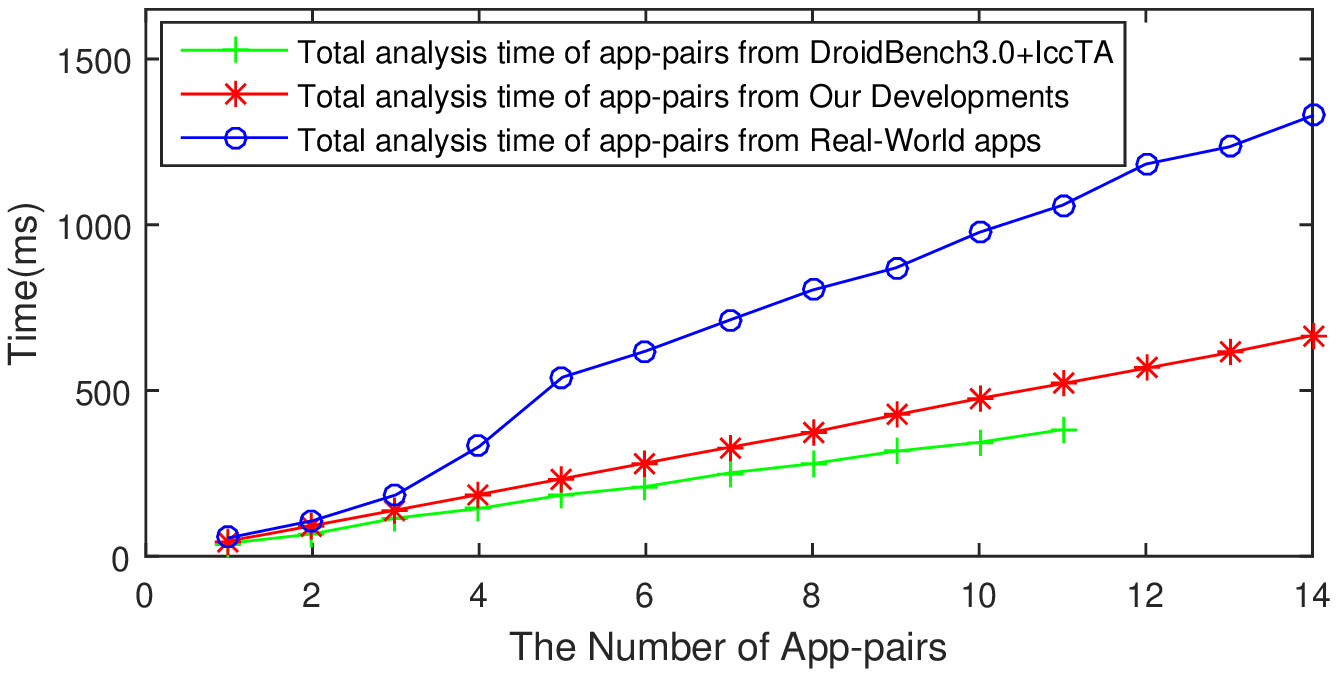} \\
\end{minipage}}%
\subfigure[Single App-pair.]{
\label{total time of analysis:single} 
\begin{minipage}[b]{0.45\textwidth}
\centering
\includegraphics[scale=0.4]{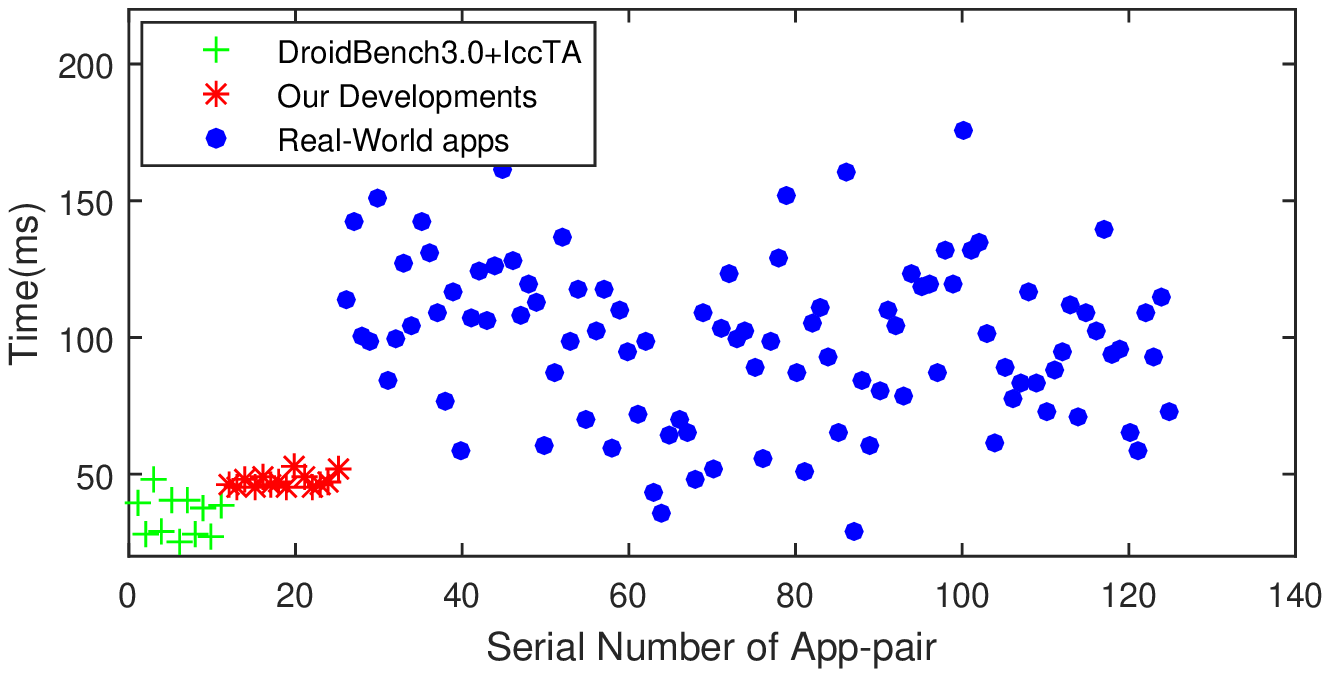} \\
\end{minipage}}
\caption{The time cost of \emph{Analyzer} under various app-pairs.}
\label{total time of analysis} 
\end{figure*}

\subsubsection{Analyzer Performance}
Figure \ref{total time of analysis:varying number} depicts the total time cost of analyzing multiple app-pairs with our \emph{Analyzer} as the number of app-pairs increases. Obviously, the entire time cost increases almost linearly along with the number of app-pairs. Thanks to the model deflation technology aforementioned, the average time cost is less than 100ms per app-pair. The time cost of app-pairs in Real-World rises sharply when the number of app-pairs reaches five, owing to some large size apps that generate more complex models. While most of the app-pairs from DroidBench3.0+IccTA and Our Development only have simple structures and functions used for experimental purposes.

To investigate the influence of app size in-depth, we analyze a variety of app size dependent factors that affect the time cost, such as the number of models, the size of log files, the amount of codes, and so on. We find that the number of models is the most influential factor. In this regard, we perform an experiment which takes about 13.7s to analyze a log file containing about 200 models (including attack and normal models), indicating an average time cost of 68.5ms per model.

Figure \ref{total time of analysis:single} further illustrates the average time cost of analyzing the logs of 140 different app-pairs with the \emph{Analyzer} in twenty runs. These 140 different app-pairs are randomly chosen from the datasets. The $x$ axis is the serial number identifying every app-pair. The $y$ axis shows the average analysis time cost of every app-pair. In consistent with Figure \ref{total time of analysis}, the time cost of the DroidBench3.0+IccTA and Our Developments concentrates on the lower time region less than 60ms. In contrast, the time cost of analyzing the logs of the app-pairs from Real-World dataset is much longer than others due to the large number of models incurred by their sophisticated functions. Therefore, by summing the runtime cost of the four parts of the \emph{Monitor} in Figure \ref{monitor runtime cost} and the time cost of single app-pair in \emph{Analyzer} in Figure \ref{total time of analysis}, the overall time overhead for processing single app-pair is less than 200ms.


\section{Discussion}
\label{discussion}
Although we made the first attempt to identify the ICC attacks by migrating TaintDroid, there are also limitations.

Firstly, to best suit our requirements, we have revised some functions and classes of TaintDroid at method-level and file-level for overcoming the challenge of ensuring the efficiency of tracking tainted data in intent communications.

Secondly, there is a necessary ongoing improvement for SIAT. Since the TaintDroid supports up to Android 4.3, the current version of SIAT is based on AOSP version 4.3. Nevertheless, We have checked and confirmed that over 96\% of apps that we have collected randomly from the Google Play market still can run on Android 4.3 with SIAT. One essential next effort for us is to migrate the TaintDroid from the Dalvik VM based Android system to the Runtime based Android system, to retain the functionality of SIAT for future use. Also, we can migrate the state-of-the-art fine-grained taint analysis tool for Android Runtime analysis, e.g., TaintMan \cite{you2017taintman} and TaintART \cite{sun2016taintart}, to intent communications and then integrate it into SIAT via the intent primitives. On the other hand, neither the run-time permission control model built since Android 6.0, nor the increased number of Android permissions, e.g., about 104 permissions in Android 4.3, about 163 permissions in Android 10, has affections on the implementation of SIAT, owing to its adaptivity and extendibility.


\section{Conclusion}
\label{conclusion}

In this paper, we present the design and implementation of the SIAT, which provides real-time systematic tracking of privacy-sensitive data and attack visibility into how the collaborative malicious behaviors take place via intent, based on its two crucial modules: \emph{Monitor} and \emph{Analyzer}. SIAT makes the first attempt to revise the TaintDroid both at method-level and file-level, to migrate it to the app-pair ICC paths identification both at the data flow and control flow. Compared to state-of-the-art approaches, the SIAT can achieve about 25\%$\sim$200\% accuracy improvements with 1.0 precision and 0.98 recall at the cost of negligible runtime overhead.



\ifCLASSOPTIONcompsoc
  \section*{Acknowledgments}
\else
  \section*{Acknowledgment}
\fi
This work is partially supported by the National Natural Science Foundation of China under Grant No.61872130, 61872133, and 61632009; Science and Technology Project of Hunan Provincial Department of Communications under Grant No.201928; Key R $\&$ D Projects in Changsha under Grant No.kq1907103.

\vspace{-0.1in}

\end{document}